\documentstyle[12pt,aaspp4]{article}

     \newcommand{\lya}{L$\alpha$}

     \slugcomment{accepted by ApJ, 15-Jan-1997}
     \begin{document}

     \title{The High Chromospheres of the Late A Stars\footnotemark}

     \author{Theodore Simon}
     \affil{Institute for Astronomy, University of Hawaii, 2680
     Woodlawn Drive, Honolulu, HI 96822}

     \author{Wayne B. Landsman}
     \affil{Hughes STX Corp., NASA/GSFC, Code 681, Greenbelt, MD
     20771}

     \footnotetext{Based on observations with the NASA/ESA {\it Hubble
     Space Telescope (HST)}, which is operated by the Space Telescope
     Science Institute under NASA contract NAS5-26555 to the
     Association of  Universities for Research in Astronomy, Inc.}

     \begin{abstract}
     
     We report the detection of \ion{N}{5} 1239\AA\ transition region emission
     in {\it HST}/GHRS spectra of the A7 V  stars, $\alpha$ Aql
     and $\alpha$ Cep.  Our observations provide the first direct evidence
     of $1-3 \times 10^5$ K material in the atmospheres of normal A-type
     stars.  For both stars, and for the mid-A--type star $\tau^3$~Eri, we
     also report the detection of chromospheric emission in the \ion{Si}{3}
     1206\AA\ line.  At a \bv color of 0.16 and an effective temperature of
     $\sim8200$ K, $\tau^3$~Eri becomes the hottest main sequence star known
     to have a chromosphere and thus an outer convection zone.
     We see no firm evidence that the \ion{Si}{3} line surface fluxes of
     the A stars are any lower than those of moderately active, solar-type,
     G and K stars.  This contrasts sharply with their coronal X-ray emission,
     which is $>100$ times weaker than that of the later-type
     stars.  Given the strength of the \ion{N}{5} emission observed here,
     it now appears unlikely that the X-ray faintness of the A stars is
     due to their forming very cool, $\leq$1 MK coronae.  An alternative
     explanation in terms of mass loss in coronal winds remains a 
     possibility, though we conclude from moderate resolution spectra of
     the \ion{Si}{3} lines that such winds, if they exist, do not
     penetrate into the chromospheric \ion{Si}{3}--forming layers of the
     star, since the profiles of these lines are {\it not} blueshifted,
     and may well be redshifted with respect to the star.
                      
     \end{abstract}


     \section{Introduction}

     In this paper we address a long-standing question in stellar
     structure as to the locus for the onset of subsurface convection
     zones in main sequence stars.  Thirty years ago, the changeover 
     from radiative to convective envelopes was presumed to occur
     in the close vicinity of spectral type F5 (e.g., Wilson
     1966, Kraft 1967), an idea that was supported by optical
     observations of the time, which showed an absence of
     chromospheric \ion{Ca}{2} emission reversals in the spectra of
     earlier stars and a steep falloff in the axial rotation rates of
     later stars (the direct result, it was reasoned, of  magnetic
     braking and angular momentum loss due to the onset of coronal
     winds). In later years, however, improved stellar structure
     models as well as X-ray and ultraviolet (UV) observations
     from space have shown that coronae, chromospheres, and thus
     convection zones, must be present in stars that are located well
     above the F5 boundary line.

     UV spectra from the {\it International Ultraviolet Explorer
     (IUE)} telescope, for example, trace chromospheric emission
     in \lya\ and in the 1335 \AA\ lines of \ion{C}{2} along
     the main sequence into the late-A stars, near \bv = 0.20 (Simon
     \& Landsman 1991; Marilli et al. 1992; Landsman \& Simon 1993),
     while a number of A stars have been detected in X rays by the
     {\it Einstein Observatory} or by {\it ROSAT} (Schmitt et al. 
     1985; Simon, Drake, \& Kim 1995), including a handful of
     A0 stars with $\bv \approx 0.0$.  The
     X-rays of the cooler A-type stars are believed to be coronal in
     origin.  Those of the hotter stars are thought to come not
     from the A stars themselves, but from hidden late-type binary
     companions or neighboring stars that lie within the X-ray beam.
     Only rarely can this ambiguity be resolved (e.g, Schmitt \& 
     K\"urster 1993).  UV spectroscopy largely avoids this difficulty
     with the imagery, since the high temperature UV lines
     of a bright A star are expected to outshine those of virtually
     any low mass companion. On the other hand, the detection of
     chromospheric emission against the bright photosphere of an A
     star poses challenges of its own, demanding moderate-to-high
     spectral resolution, minimal scattered light,  and exceptional
     signal-to-noise ratio.  Such requirements were often beyond
     the capabilities of {\it IUE}, but can now be met routinely by
     instruments aboard {\it HST}, such as the Goddard High Resolution
     Spectrograph (GHRS).

     In previous work with the GHRS (Simon, Landsman, \& Gilliland
     1994, hereafter SLG), we obtained high quality moderate
     resolution spectra of the \ion{C}{2} 1335 \AA\ lines for 8 
     mid- to late-A stars. Only in the case of the A7 V star, $\alpha$~Aql
     (Altair), did we find evidence for chromospheric emission.
     We chose the \ion{C}{2} lines for this earlier study because
     they are among the brightest features in the UV spectra of
     G--M stars; moreover, the photosphere of an A star is much
     fainter at 1335 \AA\ than it is at longer wavelengths, where the 
     increasing brightness of the background can render even strong lines, 
     such as \ion{C}{4} 1550 \AA, completely invisible (Simon \& Landsman
     1991). The strongest A star chromospheric line accessible to the
     GHRS is \lya.  This line suffers the dual disadvantage
     of  contamination by scattered light in the Earth's atmosphere
     and attenuation by interstellar \ion{H}{1} along the line of
     sight to the star.  Nonetheless, even at low resolution and low
     S/N, {\it IUE} spectra have yielded some detections among the
     late-A and early-F stars (Landsman \& Simon 1993).  Accordingly,
     with this {\it IUE} work in mind, we have
     initiated a program of  GHRS \lya\ spectroscopy for a modest
     sample of bright A stars.  Our goal was to establish more
     stringent constraints on the locus for the onset of main sequence
     convection zones.  In this article, we report the initial
     results of our observations, which include the first
     detection of  chromospheric emission in a mid-A type star, 
     as well as the first detections of $\sim10^5$ K \ion{N}{5}
     1239\AA\ transition region emission in the spectra of two late
     A-type stars.

     \section{Observations}

     We acquired GHRS spectra of the 1200--1230 \AA\ region at
     intermediate spectral resolution with the G140M grating and the
     small science aperture for seven mid- to late-A type stars. Two
     of the principal chromospheric lines located within this interval 
     are \lya\ $\lambda$1215.67 and \ion{Si}{3} 1206.51 \AA.  The latter
     is situated within the very dark, broad wings of the photospheric
     \lya\ line, which serves to enhance the contrast of the emission
     feature against the stellar background on which it lies.  We 
     summarize the results here for four stars ($\tau^3$ Eri, 
     Altair, $\alpha$ Cep, and $\alpha$ Hyi), but we postpone
     a discussion of the less well exposed spectra of three remaining 
     stars ($\alpha$ Pic, $\iota$ UMa, and $\alpha$ Cir) to a later date.  
     For two stars, Altair and $\alpha$ Cep, we obtained
     large aperture G140M spectra, centered on the \ion{Si}{4} 
     $\lambda\lambda$1393.76, 1402.77 lines.  These spectra cover the 
     wavelength range from 1385 \AA\ to 1415 \AA.  Photospheric
     absorption lines dominate the latter pair of spectra, and no
     transition region emission is evident in either one.  We observed
     the same stars, i.e., Altair and $\alpha$~Cep, with the G140L
     grating and the large science aperture, obtaining coverage of the
     1180--1440 \AA\ region at low spectral resolution.

     The above mentioned observations were carried out in the standard
     operational modes, as described by the GHRS
     {\it Instrument Handbook} (Soderblom et al. 1995).  The relevant
     procedures include sub-stepping the spectrum by fractional
     diodes of the Digicon detector to improve the sampling, moving
     the spectrum by integral numbers of diodes to compensate for 
     detector defects, and multi-diode rotations of the
     grating to facilitate the removal of fixed pattern
     noise.  To calibrate the wavelength scales of our small aperture
     spectra, we obtained scans of an onboard Pt-Ne hollow-cathode lamp.
     The wavelength scales of the large aperture spectra were
     verified and corrected via the SPYBAL spectra that accompany each
     science observation (cf. Soderblom et al. 1995).

     The data were analyzed within STSDAS/IRAF, with minor post-processing
     in IDL.  In the way of an initial overview, Figure 1 shows our low
     resolution spectra of Altair and $\alpha$ Cep. In addition to \lya,
     several features, which cannot be seen even in the best images from
     {\it IUE} because of the high scattered light levels and low S/N
     ratio, are observed here for the very first time in A-star spectra.
     These include \ion{C}{3} 1176 \AA, \ion{Si}{3} 1206 \AA,
     and \ion{N}{5} 1239 \AA. The temperatures at which these lines form
     range from $3 \times 10^4$ K to $2 \times 10^5$ K.  The presence of
     \ion{N}{5} is particularly noteworthy, as it represents the highest
     excitation stage recorded to date in the UV for any normal A star.
     At longer wavelengths, there appears to be an emission feature near
     1400 \AA.  In our G140M spectra, the same peak is located at 
     1394.34 \AA, much too long a wavelength for it to be the transition 
     region line of \ion{Si}{4}.  According to an ATLAS9 model, this 
     part of the photospheric spectrum is dominated by absorption lines 
     of \ion{Fe}{2}.  We find, by numerically broadening the model flux
     distribution to $v \sin i = 240$ km s$^{-1}$, that there is a reasonable
     match to the overall shape of our G140L observation, including the 
     observed ``emission peak.''  We take this as an indication that the
     aforementioned peak is a window of reduced line blanketing rather than
     a strongly wavelength-shifted transition region emission line.

     The \ion{Si}{3} line profiles from three of our G140M small
     aperture observations are shown in Figure 2.  The spectrum of
     $\tau^3$ Eri is weak, but presents what we regard as marginal 
     evidence for an emission line.  A much deeper exposure is 
     needed for confirmation. In the same spectrum, we see relatively
     intense emission in \lya, at a level of $2.9 \times 10^{-13}$
     erg cm$^{-2}$ s$^{-1}$, and so there is no question but that
     $\tau^3$ Eri has a chromosphere.  At \bv = 0.16, $\tau^3$~Eri
     becomes the earliest main sequence A star known to possess a
     chromosphere.

     The broad, flat-topped shapes of the \ion{Si}{3} spectra for Altair
     and $\alpha$ Cep, also shown in Figure 2, suggest considerable rotational
     broadening.  However, in both cases the FWHM and FWZI imply $v
     \sin i \gtrsim 350$ km s$^{-1}$, which is substantially greater than
     the rotation rates of $\sim$240 km s$^{-1}$ customarily cited for
     these stars. Other examples of excess broadening in GHRS spectra
     of  later-type stars are discussed by Ayres et al. (1996).  These
     authors offer several possible explanations for this effect,
     but find no entirely satisfactory answer.  The wavelength
     centroids of the \ion{Si}{3} lines of Altair and $\alpha$ Cep are
     centered on the radial velocities of the stars, or may be
     slightly offset by $+7\pm7$ km s$^{-1}$.  Similar redshifts
     are observed for other late-type stars (Ayres et al. 1996).
     Our measurements of Altair and $\alpha$~Cep appear to rule out
     any possible wind outflow at chromospheric heights for both stars
     (see SLG), since if these lines were produced in a moderate-to-high
     velocity wind we would expect them to be appreciably
     blueshifted.  Over the course of  an {\it HST} orbit, neither
     star showed significant variability in \ion{Si}{3}.  Moreover,
     the G140M and G140L \ion{Si}{3} line fluxes agree to within 4\%
     for both stars.  This sets limits on any longer-term (6--8 months)
     spectral variability, as well as on the extent of possible  
     light losses and flux calibration uncertainties in our small 
     aperture spectra.

     Table 1 summarizes the new UV line fluxes, along with revised 
     \ion{C}{2} fluxes for Altair and $\alpha$ Cep.  The latter are 
     based on the earlier study by SLG, except that we assume 
     a brightness ratio of 8:1 in accordance with the relative strengths
     of the chromospheric lines listed in Table 1, instead of the 22:1 ratio
     adopted by SLG from consideration of low S/N spectra of \lya\ from
     {\it IUE}.  Our revised \ion{C}{2} fluxes for Altair and $\alpha$
     Cep are in close agreement with, and only 18\% higher and 18\%
     lower, respectively, than those derived by Walter, Matthews, \&
     Linsky (1995) by means of an entirely different approach.

     \section{Discussion}

     To gauge the intrinsic activity of the A stars, we use the 
     Barnes-Evans (1976) relation to convert the apparent fluxes in
     Table 1, $f_{\lambda}$, to surface fluxes at the star, $F_{\lambda}$.
     The results are in Table 2, along with comparable numbers for
     X-ray emission.  For comparison, we also tabulate surface fluxes for
     representative G and K stars.  These values derive mainly from GHRS
     spectra (e.g., Ayres et al. 1997) or, in some cases, from compilations
     of {\it IUE} and {\it ROSAT} data that have appeared in the literature
     (e.g., Ayres et al. 1995).  

     In terms of their chromospheric or transition region fluxes, the A
     stars resemble moderately active late-type dwarfs or giants
     (e.g., $\epsilon$ Eri or $\beta$ Cet), but they are (to no surprise)
     much less active than the rapidly-rotating Hertzsprung Gap giants 
     (e.g., 31 Com) and the tidally locked RS CVn binaries (e.g., V711 Tau).
     In coronal X rays, the A stars are comparable to middle aged stars like
     the Sun and $\alpha^1$ Cen.  They are substantially less active than
     prototypical, rapidly rotating, young dwarf stars like $\chi^1$ Ori.
     Nor are they as active as any of the late-type giants.

     These trends are more clearly illustrated in the H--R bubble
     diagram shown in Figure 3.  In this figure, we scale the  diameter of
     the circle for each star to the common logarithm of the
     normalized flux value, $F_{\lambda}/\sigma T_{\rm eff}^4$. Turning to 
     the top panel, we find no evidence in the \ion{Si}{3} line for a decline
     in chromospheric activity along the main sequence, even among the middle
     A stars (that is, for effective temperatures as high as 8200 K) if
     one includes the uncertain measurement of $\tau^3$~Eri. 
     By contrast, the coronal X-ray fluxes in the lower panel of the
     figure show an enormous decline from one side of the diagram to the
     other, with the main sequence and more evolved early type stars to the
     left being considerably fainter than the late-type stars on the right. 
     Attention was drawn to the lower coronal/chromospheric flux ratios
     among the early type stars by Simon \& Drake (1989), who ascribed the
     effect to mass loss via coronal winds; a more recent discussion
     by Ayres et al. (1997) concludes that the origin of these ``X-ray
     deficits''\/ may be related to the possible existence of very long
     coronal loops, but otherwise the trend remains unexplained.
     At intermediate,
     transition region temperatures, a similar trend may be appearing in the 
     \ion{N}{5}/\ion{Si}{3} and \ion{N}{5}/\ion{C}{2} ratios, each being a
     factor of $2-3\times$ smaller among the A stars than among the G--K
     stars.  However, given the lack of truly adequate numbers of high
     quality spectra of \ion{N}{5} for stars of all MK classes, this
     possibility requires further study with the {\it HST}.

     Continuing the tendency toward a broad dispersion in line strengths
     among  the early F stars that was noted in our previous work (Simon 
     \& Landsman 1991; Landsman \& Simon 1993), the surface fluxes of the
     A stars in Table 2 indicate that Altair is at least a factor of 2 more
     active than $\alpha$ Cep over the entire range of temperatures
     from chromospheric to coronal, despite the otherwise close similarities
     of these two stars ($\alpha$ Cep has the slightly lower surface gravity).
     Our spectra also largely put to rest the notion that
     the weak X-ray emission of these particular stars might be linked to
     a shift in their peak coronal heating below $\sim1$ million K and to
     the formation of very cool coronae. The observed \ion{N}{5} line
     strengths, and their ratios to fluxes of cooler chromospheric lines, 
     make that appear extremely unlikely.   However, only measurements of lines
     formed at 0.5--1 million K can answer this question for certain, and
     the needed observations await the launch of future space missions.

     Support for this work was provided by NASA through grants
     GO-6083.01-94 and GO-6446.01-95 from the Space Telescope Science
     Institute, which is operated by the Association of Universities
     for Research in Astronomy, Inc., under NASA contract NAS5-26555.        


     \newpage

\begin{deluxetable}{lrlcrrrr}
\tablecolumns{8}
\tablewidth{0pc}
\tablenum{1}
\tablecaption{GHRS Emission Line Fluxes\tablenotemark{a}}
\tablehead{
\colhead{Star}		                  & \colhead{  HD}		        &
\colhead{Sp. Ty.}                         & \colhead{ \bv}			&
\colhead{$f_{1176}$}			  & \colhead{$f_{1206}$} 		&
\colhead{$f_{1239}$\tablenotemark{b}}	  & \colhead{$f_{1335}$}     
}
\startdata   
$\tau^3$ Eri	&  18978 &  A4 IV      &  0.16  &  \nodata  &   10  &  \nodata   &  \nodata  		    \nl
$\alpha$ Aql	& 187642 &  A7 V       &  0.22  &     160   &  175  & 21\phd\phn &  540\tablenotemark{c}     \nl
$\alpha$ Cep    & 203280 &  A7 V$^{+}$ &  0.22  &      18   &   23  &      2.2   &   68\tablenotemark{c}     \nl
$\alpha$ Hyi    &  12311 &  F0 V       &  0.28  &  \nodata  &   39  &  \nodata   &   76\tablenotemark{d}     \nl 
\tablenotetext{a}{As observed at Earth.  In units of 10$^{-14}$ erg cm$^{-2}$ s$^{-1}$.}
\tablenotetext{b}{Includes only the stronger line of the doublet.  The second 
line, at 1242.80 \AA, is weak in the G140L spectrum of Altair and absent from the
spectrum of $\alpha$ Cep.  We believe this is due to absorption lines in the
underlying spectrum of the photosphere.}
\tablenotetext{c}{Assuming an $\alpha$ Aql/$\alpha$ Cep \ion{C}{2} brightness ratio of 8:1.  See text and SLG.}
\tablenotetext{d}{From {\it IUE} spectrum, Simon \& Landsman 1991.}
\enddata
\end{deluxetable}
\newpage

\begin{deluxetable}{lrrrrrrr}
\tablecolumns{8}
\tablewidth{0pc}
\tablenum{2}
\tablecaption{Emission Line Surface Fluxes\tablenotemark{a}}
\tablehead{
\colhead{Star}		                  & \colhead{ \bv}		        &
\colhead{$M_v$}                           & \colhead{$T_{\rm eff}$}		&
\colhead{L/L$_{\sun}$}			  & \colhead{$F_{1206}$} 		&
\colhead{$F_{1239}$}			  & \colhead{$F_X$}     
}
\startdata   
$\tau^3$ Eri	&  0.16  &  2.63   &  8210  &   6.8  &  47\phd\phn &  \nodata       &   \nodata      	    \nl
$\alpha$ Pic	&  0.21  &  1.85   &  7580  &  13.6  &  \nodata    &  \nodata       &   43\phn              \nl
$\alpha$ Aql    &  0.22  &  2.30   &  7890  &   9.1  &  37\phd\phn &   4.4\phn      &   13\phn              \nl
$\alpha$ Cep    &  0.22  &  1.60   &  7720  &  17.3  &  17\phd\phn &   1.6\phn      &   5.6      	    \nl
$\alpha$ Hyi    &  0.28  &  1.27   &  7190  &  24.0  &  34\phd\phn &  \nodata       &   42\phn              \nl
$\beta$ Cas     &  0.34  &  1.56   &  6870  &  18.5  &  59\phd\phn &   7.7\phn      &   47\phn              \nl
$\alpha$ C\/Mi  &  0.42  &  2.71   &  6270  &   6.6  &  28\phd\phn &   5.4\phn      &   60\phn	   	    \nl
$\chi^1$ Ori    &  0.59  &  4.50   &  5800  &   1.3  &  \nodata    &   3.9\phn      &   1600\phn	    \nl
$\alpha$ Aur Ab &  0.60  &  0.16   &  5580  &  72.5  & 120\phd\phn & 23\phd\phn\phn &   200\phn		    \nl
$\beta$ Hyi	&  0.62  &  3.81   &  5860  &   2.5  &  \nodata    &   0.80         &   6.3	      	    \nl
Quiet Sun	&  0.63  &  4.84   &  5771  &   1.0  &  9.5        &   0.64         &   20\phn	     	    \nl
31 Com		&  0.67  &  0.15   &  5570  &  73.0  & 150\phd\phn & 27\phd\phn\phn &   1300\phn	    \nl
$\kappa$ Cet	&  0.68  &  5.01   &  5510  &   0.9  &  \nodata    &   6.2\phn      &   950\phn             \nl
$\psi^3$ Psc   & 0.69 & $\sim$0.15 &  5520  &  73.0  & 110\phd\phn & 18\phd\phn\phn &   180\phn	            \nl
$\alpha^1$ Cen  &  0.71  &  4.37   &  5540  &   1.5  &  5.5        &   1.0\phn      &   8.8       	    \nl
$\epsilon$ Eri  &  0.88  &  6.14   &  4940  &   0.4  &  19\phd\phn &   4.9\phn      &   610\phn	 	    \nl
$\alpha^2$ Cen  &  0.88  &  5.71   &  5030  &   0.5  &  5.3        &   0.54         &   46\phn     	    \nl
$\alpha$ Aur Aa &  0.90  &  0.31   &  5190  &  73.2  &  15\phd\phn &   6.4\phn      &   150\phn		    \nl
V711 Tau B\tablenotemark{b}  &  0.99  & 3.54 & 4860  &  4.1 &  280\phd\phn  & 88\phd\phn\phn & 38000\phn    \nl
$\beta$ Cet	&  1.01  &  0.95   &  4970  &  43.6  &  10\phd\phn &   6.1\phn      &   130\phn		    \nl
$\lambda$ And	&  1.02  &  2.01   &  4750  &  17.3  &  46\phd\phn & 22\phd\phn\phn &   1700\phn	    \nl 
\tablenotetext{a}{In units of 10$^{3}$ erg cm$^{-2}$ s$^{-1}$}
\tablenotetext{b}{Entries are for the active K1 IV secondary star of this spectroscopic binary.}
\enddata
\end{deluxetable}

     \newpage
     \begin{figure}
     \plotone{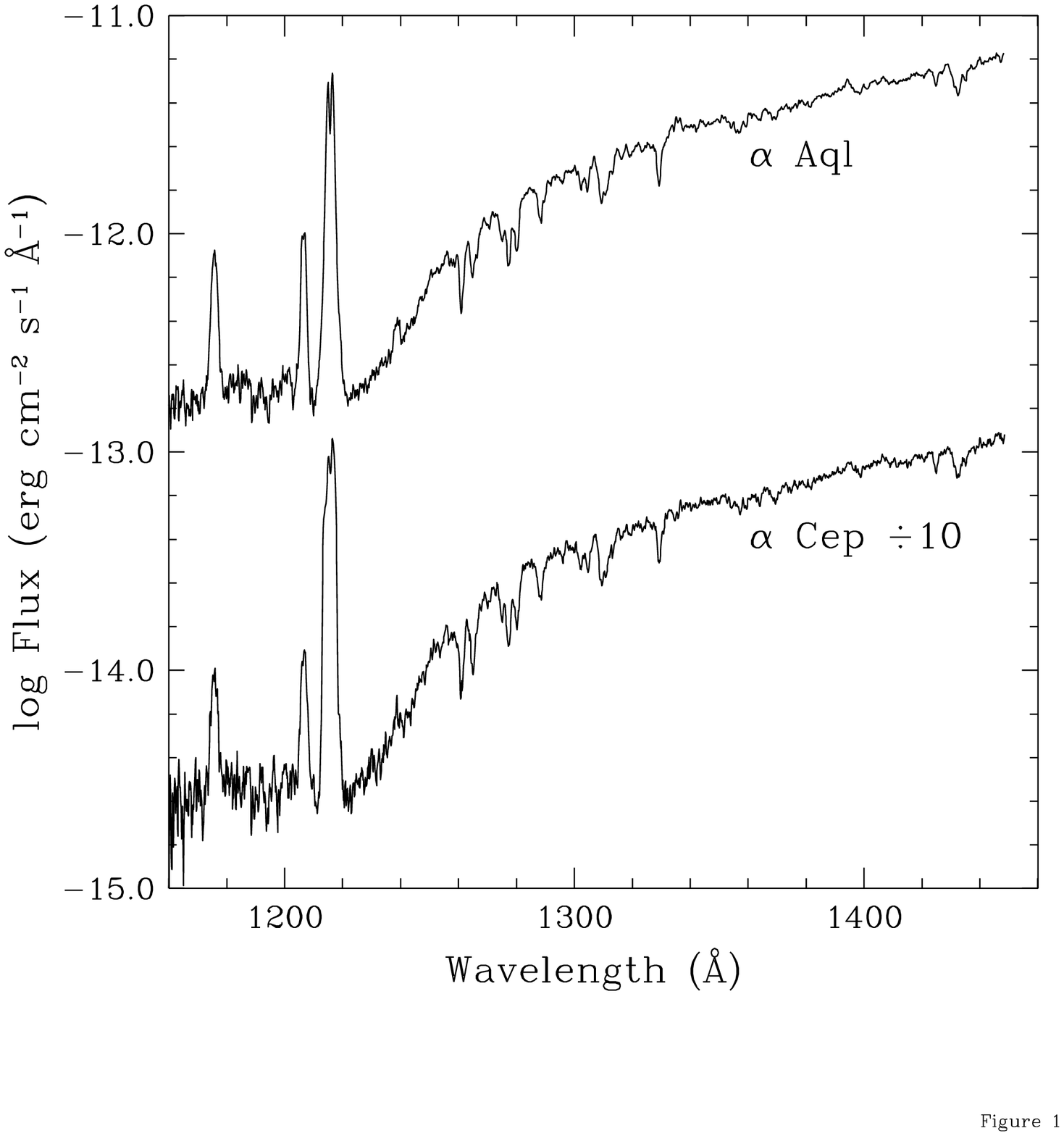}
     \caption {Low resolution GHRS G140L spectra of Altair and
     $\alpha$ Cep, showing stellar emission in C III, Si III,
     H I, and N V.}
     \end{figure}

     \begin{figure}
     \plotone{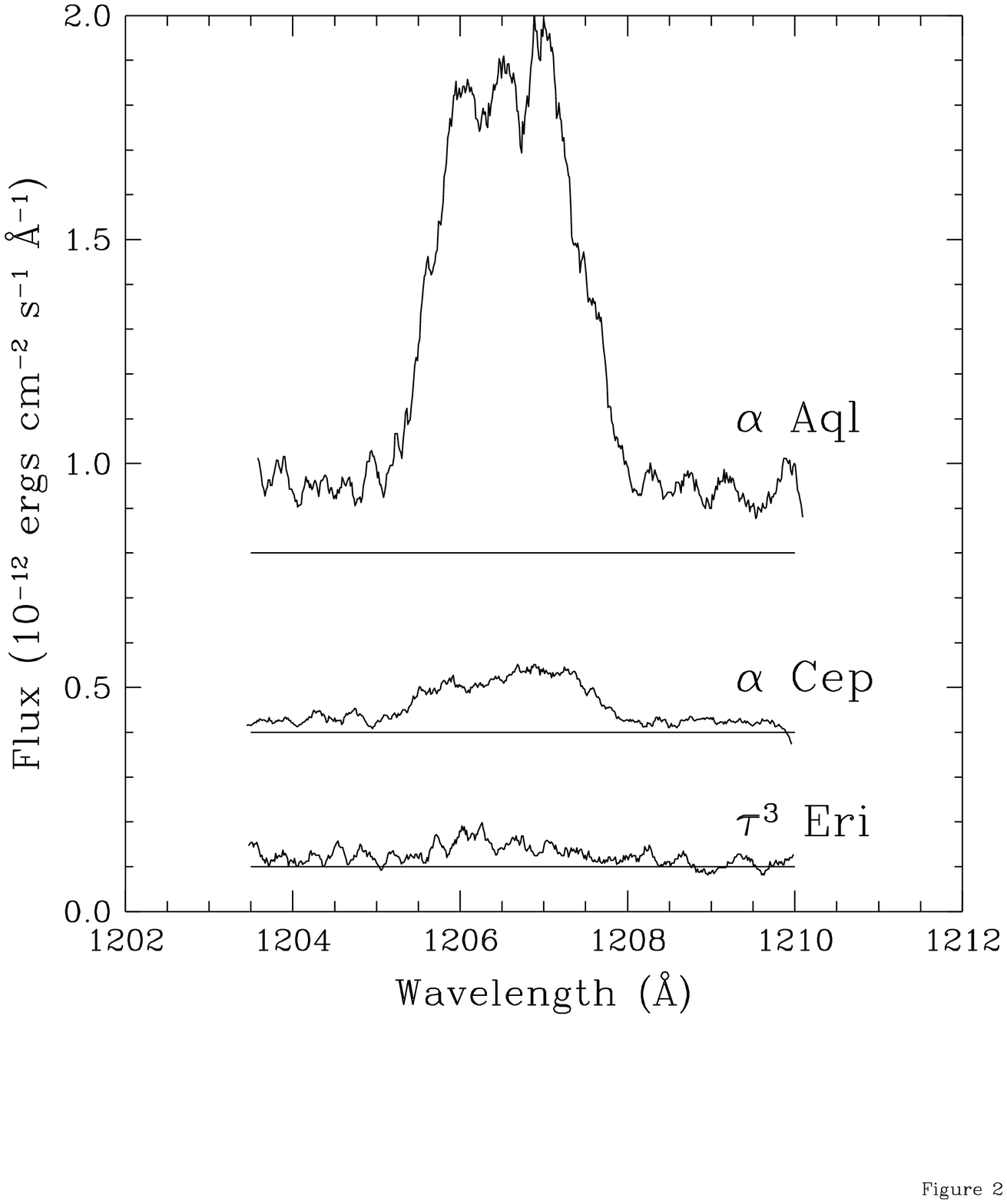}
     \caption {GHRS G140M small aperture spectra of Si III for
     three A stars.  The data have been smoothed with a 3-diode
     running average.  The individual profiles have been offset
     vertically, as indicated by the horizontal baselines.}
     \end{figure}

     \begin{figure}
     \plotone{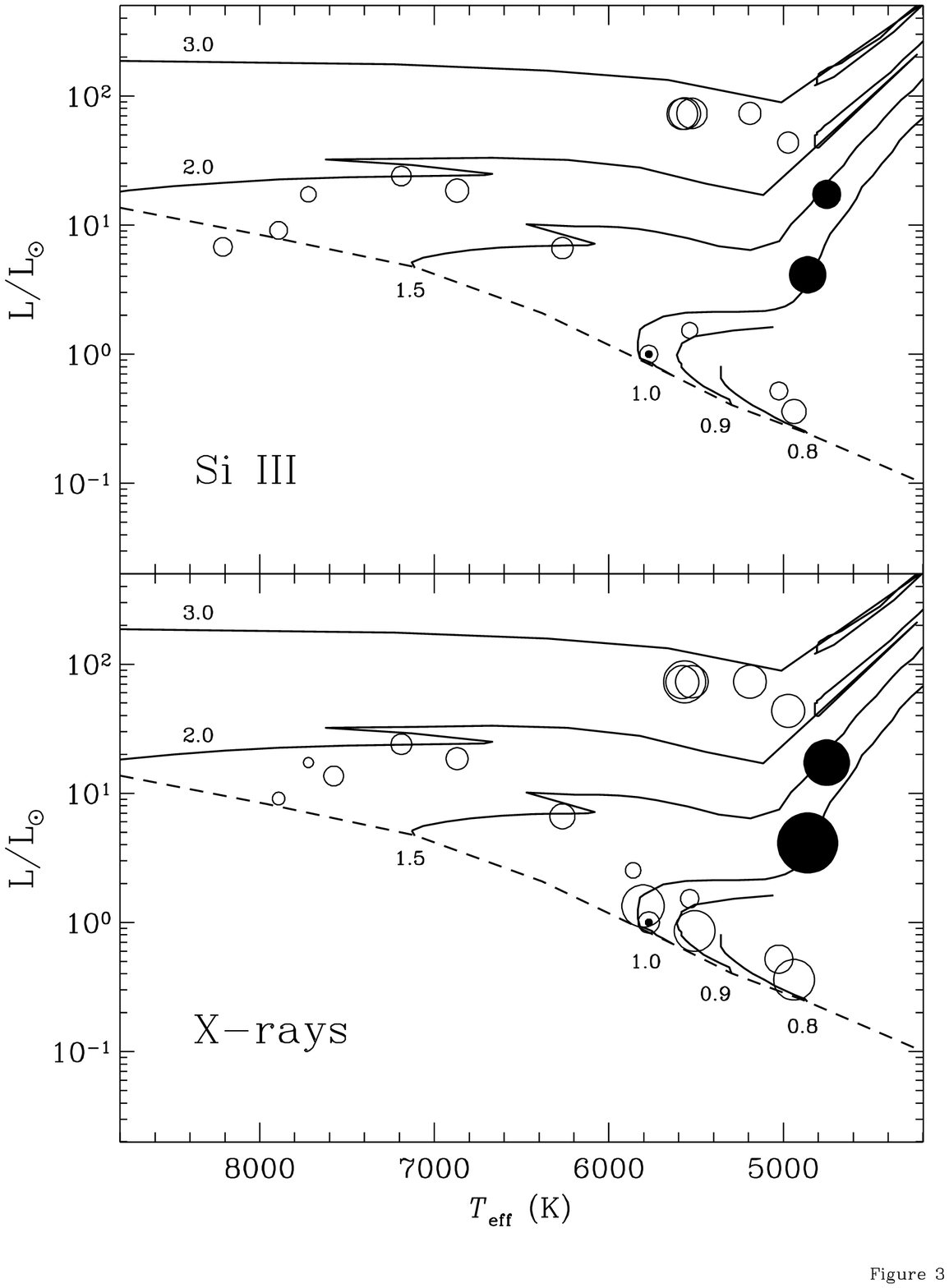}
     \figcaption {An H--R bubble diagram for the stars in Table 2.
     The dimensions of the circles are proportional to the log of
     the normalized flux, $\log(F_{\lambda}/\sigma T_{\rm eff}^4)$,
     for the Si III emission line ({\it top panel}) and for X-ray
     emission ({\it bottom panel}).  The two large filled circles
     denote RS CVn binaries.  The small dotted circle represents
     the Quiet Sun. The evolutionary tracks, with masses labeled 
     in solar units, are from Schaller et al. 1992.}
     \end{figure}

     \end{document}